\documentclass{scrartcl}
\usepackage{amsmath}
\usepackage{amscd}
\usepackage{graphicx}
\usepackage[utf8]{inputenc}
\usepackage{cite}

\author{J. A. S. Lima \footnote{limajas@astro.iag.usp.br}\\
	Departamento de Astronomia, Universidade de S\~ao Paulo, \\05508-900 S\~ao Paulo, SP, Brazil, \\
	\and
	F. D. Sasse \footnote{fernando.sasse@udesc.br}\\
	Departamento de Matem\'atica, \\Universidade do Estado de Santa Catarina, \\89219-710 Joinville, SC, Brazil}

\title{Can Lorentz transformations be determined by the null Michelson-Morley result?\footnote{Published on Revista Brasileira de Ensino de F\'{\i}sica \textbf{39}, 3, e3304 (2017).}}

\begin{document}
	\maketitle
	
	The so-called principle of relativity  is able to fix a general coordinate transformation which differs from the standard  Lorentzian form only by an unknown speed which cannot in principle be identified with the light speed. Based on a reanalysis the Michelson-Morley experiment using this extended transformation we show that such unknown speed is analytically determined regardless of the Maxwell equations and conceptual issues related to synchronization procedures, time and causality definitions. Such a result demonstrates in a pedagogical manner that the constancy of the speed of light does not need to be assumed as a basic postulate of the special relativity theory since it can be directly deduced from an optical experiment in combination with the principle of relativity. The approach presented here provides a simple and insightful derivation of the Lorentz transformations appropriated for an introductory special relativity theory course.\\
	\textbf{Keywords:} Michelson-Morley experiment, Lorentz transformations, principle of relativity\\

\section{Introduction}

In the standard lore of special relativity theory (SRT), the pillars of the theory  rest on two postulates originally introduced by Einstein \cite{einstein1905}, namely: (i) the principle of relativity, and (ii) the principle that states that the speed of light is independent of the velocity of the source (see ref. \cite{baierlein2006} for an explanation of why this is not the same as ``the constancy of the speed of light").  These two postulates were explicitly used by him for obtaining the so-called Lorentz transformations. Nevertheless, since the first decade after Einstein's seminal paper \cite{einstein1905}, many authors have tried to show that the second postulate is not necessary. The first attempt was made by  Ignatowski \cite{ignatowski1910} in 1910.  He replaced the second Einstein postulate by  the assumption of  isotropy and homogeneity of space, which implies linearity of the transformation equations and the reciprocity of the coordinates transformation - which means  that two inertial observers must agree with the numerical value of their relative velocities.  In 1911, Frank and  Rothe \cite{frank1911} derived the Lorentz transformations by assuming  that they form a homogeneous linear group,  the validity of reciprocity principle and the dependency of the length contraction only on the relative velocity. In 1921, Pars \cite{pars1921} derived  the Lorentz transformation, assuming homogeneity of space-time, isotropy of  space  and the reciprocity principle.  In fact, it was shown by Berzi and Gorini \cite{berzi1969} that  the principle of relativity  and spatial isotropy imply reciprocity. Levy-Leblond \cite{levy-leblond1976} has shown that  the additional hypotheses of group law and  causality are necessary. We refer to \cite{berzi1969,gorini1970,torretti1996,brown2005} for  discussions  of the necessary hypotheses, and also to Miller \cite{miller1998}, for a  complete historical account of Ignatovski's work. Pedagogical  derivations of Lorentz transformations without the second postulate can be found, for example, in 
 \cite{weinstock1965, mitvalsky1966, lee1975, lee1976,levy-leblond1976, vargas1976, Terletskii1968, rindler1977, mermin1984, schwartz1984, schwartz1985, ross1987, sexl2001, pal2003, gannett2007, pelissetto2015, drory2015}. An additional list of references regarding derivations of this kind is given by Sonego and Pin \cite{sonego2009}.

All these derivations arrive at formulas for the Lorentz transformations containing an unknown and invariant (constant) limiting speed. However,  its identification with the speed of light usually requires the invariance of electrodynamics \cite{pauli1981} or some dynamical effect \cite{frisch1963}. The main reason for so many derivations is that some authors have different opinions about what are the most fundamental  assumptions, while others present derivations that look pedagogically simpler (see the work of Llosa \cite{llosa2014} for a comprehensive review).

One modern relevance of this result lies in the fact that to study the consequences of Lorentz symmetry breaking one has to abandon or modify the principle of relativity \cite{mattingly2005,baccetti2012,liberatti2013}. Also, when taking in account theories for varying speed of light, it is important to know the origin of the terms containing the speed of light in the equations \cite{ellis2005}.

In this work, we show how the identification  of this constant speed with the speed of light could have been made in the early years of the theory of special relativity,  by applying the derived general transformations to the null results obtained in the Michelson-Morley experiment    \cite{michelson1887}. To obtain the usual Lorentz transformations, we replace the second postulate by a careful interpretation of the empirical (null) result of that optical experiment. The  approach discussed here not only establishes the speed of light as the limiting speed to be used at Lorentz transformations but also shows explicitly that the hypothesis of the existence of a luminiferous aether does not interfere with the result since it becomes irrelevant as a consequence of the null Michelson-Morley experiment.

\section{Lorentz transformations without the second postulate}
Let us now suppose that Cartesian coordinates $(t,x,y,z)$ and $(t', x',y',z')$ are associated to the inertial frames $S$ and $S'$. The frame $S'$ moves with  speed $V$ with respect to $S$, along the positive direction of $x$ and $x^{\prime}$.  We also assume that when $t=t'=0$ all spatial axes coincide.

We choose the work of Levy-Leblond \cite{levy-leblond1976} for its elegance and generality\footnote{Llosa \cite{llosa2014} presents a similar derivation that does not require the counting of parameters of the transformation group.}. Assuming validity of the principle of relativity plus the hypotheses of homogeneity of space-time, the linearity of inertial transformations,  isotropy of space and the group law, he derived a set of coordinate transformations between two inertial frames slightly more general than that proposed by Lorentz and Einstein.

 Using the standard configuration coordinates of the inertial frames $S$ and $S^{\prime}$ described above, the general  transformation set is given by:
\begin{align}
\label{eq-lorentz1}
&x'=\Gamma(\sigma)(x-Vt),\\
\label{eq-lorentz2}
&y'=y,\\
\label{eq-lorentz3}
&z'=z,\\
\label{eq-lorentz4}
&t'=\Gamma(\sigma)\left(t-\frac{Vx}{\sigma^2}\right),
\end{align}
where
\begin{equation}
\label{eq-lorentz5}
\Gamma(\sigma)=\frac{1}{\sqrt{1-V^2/\sigma^2}},
\end{equation}
and $\sigma$ is an invariant (unknown)  universal constant with dimensions of speed ($0\leq
\sigma \leq \infty$).
From now on the above set will be termed $\sigma$-Lorentz transformations since  the limiting invariant speed $\sigma$ does not need to be identified with the light speed. As one may show based on the above set, the associated direct (and inverse) $\sigma$-addition velocity law  is
given by:
\begin{align}
\label{eq-Vel1}
&v_x' = \frac{v_x - V}{1 -v_x V/\sigma^{2}},  &v_x = \frac{v_x' + V}{1 + v_x'V/\sigma^{2}},\\
\label{eq-Vel2}
&v_y' = \frac{v_y\sqrt{1-V^2/\sigma^2}}{1 -v_x V/\sigma^{2}}, &v_y = \frac{v_y'\sqrt{1-V^2/\sigma^2}}{1 +v_x' V/\sigma^{2}},\\
\label{eq-Vel3}
&v_z' = \frac{v_z\sqrt{1-V^2/\sigma^2}}{1 - v_x V/\sigma^{2}},  &v_z = \frac{v_z'\sqrt{1-V^2/\sigma^2}}{1 +v_x V/\sigma^{2}},
\end{align}
where $v_x',v_y',v_z'$ and $v_x,v_y,v_z$  are the Cartesian components of the  velocities in the inertial frames $S'$ and $S$, respectively. Clearly, all the above formulas have  the non-relativistic Galilean results as a limit when $V/\sigma \rightarrow 0$.

It is also worth mentioning that the above transformations are a particular case of the more general set assumed by Robertson \cite{robertson49}, from which he concluded that the three second-order optical experiments taken together, namely:
Michelson-Morley \cite{michelson1887} (1887),  Kennedy-Thorndike \cite{KT32} (1932),
and Ives-Stilwell \cite{IS38,IS41} (1938, 1941) are sufficient to single out the Lorentz transformations.

In contrast, we have found that a suitable analysis of any of the cited  optical experiments is enough to obtain the relativistic result, that is, $\sigma=c$.   For the sake of pedagogical simplicity, in what follows we choose the Michelson-Morley experiment, which  will be reanalyzed  based on the transformation set given by equations (\ref{eq-lorentz1}-\ref{eq-lorentz4}) plus the above $\sigma$-addition  velocity law. We notice that in most textbooks the Michelson-Morley experiment is analyzed only in the context of prerelativistic physics.

\section{Michelson-Morley experiment and the limiting invariant speed}
\label{sec-MM}
\begin{figure}[h]
\centering
\includegraphics[scale=0.7]{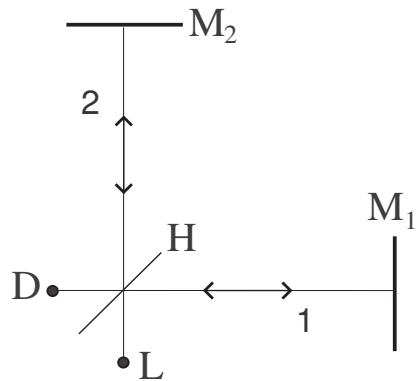}
\caption{Diagram of the Michelson-Morley experiment.}
\label{fig-mich}
\end{figure}

The Michelson-Morley experiment consists of an optical interferometer assembled on a platform
that can horizontally be rotated. A simplified diagram is shown in Figure \ref{fig-mich}.
We suppose that a beam of  light  coming from a source $L$ reaches a semi-transparent mirror $H$, which divides the original
beam into two other  beams that propagate along the orthogonal arms 1 and 2.  On the extremity
of each arm, the beams are reflected by mirrors $M_1$ and $M_2$, returning to the semi-transparent mirror $H$. A fringe shift associated with the interference (caused by the difference in the optical path) is registered at the detector $D$. The experiment aims  to observe the dependence of the  fringe shift with the spatial orientation of the apparatus.  A non-null result would indicate an anisotropy of the speed of light in the frame of the apparatus \cite{brown2005}.

Let us  suppose that the whole interferometer is at rest in the lab frame $S'$ which is moving with speed $V$ relative to an inertial (aether) frame $S$ where, by hypothesis, the speed of light $c$ is isotropic but not necessarily equal to $\sigma$. We assume that light moves in $S'$  with velocity  $\mathbf{c}'$, whose magnitude is not  necessarily isotropic, and is related to $\mathbf{c}$ by the  $\sigma$-Lorentz velocity  transformation law  given by equations (\ref{eq-Vel1}ef{eq-Vel3}).

We suppose  that the apparatus is configured in such a way that the arm 1 is aligned to this constant relative speed $V$ during the experiment. Our task now is to calculate the travel time difference between the beams along both arms in $S'$.
We assume that the arm 1  of the interferometer is directed along the $x$ axis and the arm 2 is directed along the $y$ axis so that $c_z(1)=c_z(2)=0$. Further, as one may check, the components $c_x(1)$, $c_y(1)$ and $c_x(2)$, $c_y(2)$ of the light velocity  along the arms 1 and 2 as seen in the aether frame $S$ are given by:
\begin{align}
\label{eq-velS}
&c_x(1)=c,\quad c_y(1)=0,\quad c_x(2)=V,\\&c_y(2)=c\sqrt{1-\frac{V^2}{c^{2}}}.
\end{align}
Now, by using the $\sigma$-relativistic velocity transformations given by equations (\ref{eq-Vel1}-\ref{eq-Vel2}), it is readily checked
that  the components of the  light velocity in each arm calculated in the frame $S'$ are:
\begin{equation}
\label{eq-mich1}
c_x'(1)^{\pm}=\frac{c\mp V}{1\mp cV/\sigma^2}\,,\qquad c_y'(1)=0,
\end{equation}
and
\begin{equation}
\label{eq-mich2}
c_x'(2)=0\,,\qquad  c_y'(2)=\frac{c\sqrt{1-V^2/c^2}}{\sqrt{1-V^2/\sigma^{2}}},
\end{equation}
where $c_x'(1)^{\pm}$ are the speed of the beam in the positive and negative $x$ axis direction, respectively.

Let us denote by $L_1'$ and $L_2'$, the lengths of the optical paths 1 and 2, respectively, as measured in the lab frame $S'$.
From equation (\ref{eq-mich1}) we find that the time $T_1'$ for the beam 1 to make a round trip along $L_1'$ is
  \begin{align}
\label{eq-mich3a}
T_1'=&\frac{L_1'}{c_x'(1)^{+}}+\frac{L_1'}{c_x'(1)^{-}}\\=&L_1' \left(\frac{1-cV/\sigma^2}{c-V}+\frac{1+cV/\sigma^2}{c+V}\right)\\
=&\frac{2L_1'}{c}\left(\frac{1-V^2/\sigma^2}{1-V^2/c^2}\right)=
\frac{2L_1'}{c}\epsilon^2,
\end{align}
where
\begin{equation}
\label{eq-epsilon}
\epsilon:=\sqrt{\frac{1-V^2/\sigma^2}{1-V^2/c^2}}.
\end{equation}

By using equation (\ref{eq-mich2}) we also find that the time $T_2'$ for the beam 2 to make a round trip along $L_2'$ is
\begin{equation}
\label{eq-mich3b}
T_2'=\frac{2 L_2'}{c_y'(2)}=\frac{2 L_2'}{c}\frac{\sqrt{1-V^2/\sigma^2}}{ \sqrt{1-V^2/c^2}}=\frac{2L_2'}{c}\epsilon.
\end{equation}
The interference pattern, between the light beam coming out of the two optical paths  is determined by the time delay
\begin{equation}
\label{eq-mich5}
\Delta T'= T_1'-T_2'= \frac{2\epsilon}{c}(\epsilon L_1'-L_2').
 \end{equation}

When the interferometer is rotated clockwise by $90^{\circ}$, the roles
of arms 1 and 2 are interchanged and the  times to a round trip  are modified to:
\begin{equation}
\label{eq-mich6b}
\overline{T_1'}=\frac{2L_1'}{c}\epsilon,\qquad \overline{T_2'}=\frac{2L_2'}{c}\epsilon^2,
\end{equation}
while the corresponding time delay  reads:
\begin{equation}
\label{eq-mich7}
\overline{\Delta T'}= \overline{T_1'}-\overline{T_2'}=\frac{2\epsilon}{c}(L_1'-\epsilon L_2').
 \end{equation}

Due to the rotation of the apparatus, there is a net difference in the time delays associated to each angular configuration:
\begin{equation}
\label{eq-mich8}
\Delta \mathcal{T}'=\Delta T'- \overline{\Delta T'}=\frac{2(L_1'+L_2')}{c}\,\epsilon\,(\epsilon-1).
\end{equation}

The expected fringe shift after rotation of the apparatus can be written as the ratio:
\begin{equation}
\label{eq-mich9}
\Delta N \equiv \frac{\Delta \mathcal{T}'}{\mathcal{P}'},
\end{equation}
where  $\mathcal{P}'$ is the period of the wave arriving at the detector (lab frame $S'$).

In the original experiment, using multiple reflections, the total length  of the arms ($L_1'+L_2'$) was effectively increased to eleven meters, and, as such,  some fringe displacement  would  have been observed. However, if such displacement existed, it would have to be less than 0.01 of a fringe. A modern statistical analysis of the 1887 Michelson-Morley experiment shows  that no fringe shift was observed within the accuracy limit of the apparatus, i.e., $\Delta N=0$ \cite{handschy1982}.

For an arbitrary relative speed $V < \sigma$ (see equation (\ref{eq-lorentz5})), we can conclude from equation (\ref{eq-mich8}) that  the unique  solution  of equation (\ref{eq-mich9}) for $\Delta N=0$ (consistent with the null result) is $\epsilon=1$, i.e., $\sigma=c$. Thus, the set of transformations given by equations (\ref{eq-lorentz1}-\ref{eq-lorentz5}) reduces to  the standard Lorentz form. Naturally, since the invariant speed $\sigma$ (which is a consequence of the space-time isotropy) can now be identified with the light speed in the aether frame,  this result also suggests the non-existence of an aether medium itself because of the inferred invariance of the light waves speed.  Also, from equations  (\ref{eq-mich1}) and (\ref{eq-mich2}) one may also check that the null result now also implies the isotropy of the light speed in $S'$, that is,   $c_x'(1)^{\pm}=c_y'(2)=c$.

We remark that the Michelson-Morley experiment measures only the isotropy of the two-way speed of light. Therefore, its result does not depend on the synchronization procedure \cite{will1982}.

\section{Conclusions}

Standard derivations of the generalized Lorentz transformations given by equations (\ref{eq-lorentz1}-\ref{eq-lorentz5})  show that the unique free-parameter to be determined is an invariant (and unknown) maximum speed, $\sigma$. In this paper we have shown that a non-Galilean analysis of the  null Michelson-Morley experiment provides the identification  $\sigma=c$ thereby fixing the standard Lorentz transformation.  Historically,  this remarkable result could have been obtained  before the Lorentz and Einstein derivations. However, at that time,  the concepts in physics were so permeated by Newtonian ideas that any attempt to adopt the homogeneity and isotropy of the space-time as a fundamental principle and study its consequences would appear too bizarre.  These speculations should be taken with extreme caution, however, since many  dangers haunt those who venture to use plain modern knowledge to analyze the genesis of scientific theories \cite{martins2001}.

The step from the generalized Lorentz transformations to the usual ones presented here has a methodological and also a clear  pedagogical advantage for undergraduate teaching. In particular, it does not require from fresh undergraduate students a previous knowledge of Maxwell's equations. More  puzzling kinematic concepts, like the relativity of simultaneity and synchronization procedures \cite{mansouri1977, ohanian2004, martinez2005} can be postponed for a second  study  of the Lorentz transformations.

Another interesting pedagogical aspect of our complete derivation without the second postulate is that the existence of the luminiferous aether  was explicitly assumed from the very beginning, but its possible effects on the light propagation work  only to provide the expected identification of the invariant undetermined speed,  namely: $\sigma \equiv c$.
\\ \\
\textbf{Acknowledgments}\\ \\
J. A. S. Lima is partially supported by CNPq and FAPESP (Brazilian Research Agencies).\\ \\

\end{document}